 \documentclass[sort&compress,preprint,review,12pt]{elsarticle}

 \usepackage{graphicx}

\usepackage{amssymb}

\usepackage[usenames]{color}

\journal{Physics Letters A}

\begin{document}

\begin{frontmatter}



\title{Generalized synchronization of chaos for secure communication: remarkable stability to noise}


\author[SSU]{Olga~I.~Moskalenko\corref{lulu}}
\cortext[lulu]{Corresponding Author} \ead{moskalenko@nonlin.sgu.ru}
\author[SSU]{Alexey~A.~Koronovskii}
\ead{alkor@nonlin.sgu.ru}
\author[SSU]{Alexander~E.~Hramov}
\ead{aeh@nonlin.sgu.ru} \address[SSU]{Faculty of Nonlinear
Processes, Saratov State University, 83, Astrakhanskaya, Saratov,
410012, Russia}

\begin{abstract}
A new method for secure information transmission based on
generalized synchronization is proposed. The principal advantage of
it is a remarkable stability to noise. To reveal this peculiarity of
the proposed method the effectiveness of the detection of the
information signal from the transmitted one in the presence of noise
in the communication channel is examined both for the proposed
scheme and for the schemes of chaotic communication known already.
The main ideas of the proposed method are illustrated by the example
of coupled R\"ossler systems used both transmitter and receiver.
\end{abstract}

\begin{keyword}
synchronization \sep chaotic oscillators \sep dynamical system \sep
generalized synchronization \sep noise \sep chaotic communication
\sep secure information transmission

\PACS 05.45.Pq \sep 05.45.Tp \sep 05.45.Vx \sep 05.45.Xt


\end{keyword}

\end{frontmatter}

\section*{INTRODUCTION}
\label{sct:Introduction} The use of chaotic signals for information
transmission attracts great attention of modern
scientists~\cite{Parlitz:1992_ChaoticCommunication,Cuomo:1993_ChaosCommunication,Kocarev:1995_ChaoticComunication,
Peng:1996_GyperChaosSynchro,Anishchenko:1998_Communication,Anischenko:InfoProt1998,Eguia:2000,Fischer:2000_ChaosCommunication,
Rulkov:2002_ChaoticCommunication,Yuan_SecureCommunication:2005,Li_ChaoticTrans:2006,Fradkov_ChaosTrans:2006}.
Along with the theoretical studies, the chaos based systems are used
in the practical
applications~\cite{Larsen:2006_Monterej,Dmitriev:2003_IJBC_CP,Dmitriev:1995_ChaoticCommunications}.
Among of the works devoted to this subject there are a lot of papers
devoted to the secure
communication~\cite{Roy:Chaos_Nature2005,Jaeger:Science2004ChaosComm,Parlitz:1992_ChaoticCommunication,
Cuomo:1993_ChaosCommunication,Kocarev:1995_ChaoticComunication,Peng:1996_GyperChaosSynchro,
Eguia:2000,Fischer:2000_ChaosCommunication,Rulkov:2002_ChaoticCommunication,
Dmitriev:2003_ChaoticCommunic,Dmitriev:2003_IJBC_CP,Yuan_SecureCommunication:2005,Li_ChaoticTrans:2006,Fradkov_ChaosTrans:2006,
Cessac:TransSign2006,Rohde:EstSecCom2008,Materassi:TimeScalingSecure:2008}.
Several methods for secure information transmission are known. Some
of them are based on the synchronization of chaotic
oscillators~\cite{Murali:1993_SignalTransmission,Boccaletti:1997_ChaoticCommunication,
Carroll:1998_ChaoticCommunication,Dmitriev:1999_Chanel,Terry:GSchaosCom2001,Lucamarini:SecureCommunication:2005,
Xiang:daptSecCom:2006,Cruz-Hernandez:ChuaCommun2008}. Recently, the
practical implementation of the chaotic synchronization for the
secure information transmission have been
reported~\cite{ISI:000233300200045}.

One of the most important problems connected with the secure
communication based on the chaotic synchronization phenomenon is the
distortions of signals caused by the influence of noise (first of
all, in the communication channel) that results in the loss of the
transmitted information, with the ability of the secure
communication systems to be stable to the external perturbations
being limited~\cite{Li:Breaknoise:Chaos2005}. All known schemes
based on the chaotic synchronization phenomenon prove to be
characterized by the weak robustness to
noise~\cite{Dmitriev2002:ChaosCommunication,TaoYang2004:SecureCommunication,
JinFeng:ChaosBreak2008}.

In this paper we report a new method for secure information
transmission possessing remarkable stability to noise. As it would
be shown below, our scheme demonstrates the great resistance to
noise in comparison with the other ones known hitherto. Moreover, we
use the subsidiary source of noise in the proposed scheme to provide
the additional masking of the information signal. The scheme is
based on the generalized synchronization
phenomenon~\cite{Rulkov:1995_GeneralSynchro,Aeh:2005_GS:ModifiedSystem,Boccaletti:2002_ChaosSynchro},
with several other disadvantages inherent in the already known
systems for secure communications being overcome. At the same time,
as well as other communication schemes based on chaotic
synchronization it is characterized by the low rate of information
sending.

The structure of the paper is the following.
Section~\ref{sct:GSandNIS} describes a theoretical background
leading to the new method for secure information transmission. It
contains brief description of the generalized synchronization
regime, with principal advantages of its use for secure
communication being briefly discussed. In Section~\ref{sct:Method} a
new method for secure information transmission based on this type of
the synchronous chaotic system behavior is proposed. General
requirements for such secure communication scheme are considered.
Section \ref{sct:NumSim} presents results of numerical simulation of
the proposed method. In Section \ref{sct:SNR} the stability of our
communication scheme (as well as a series of another ones) to noise
is discussed. Quantitative characteristics of the efficiency of the
secure communication schemes are introduced. Our scheme is shown to
possess the remarkable stability to noise.
Section~\ref{sct:Advantages} discusses the other advantages of the
proposed secure communication scheme. The influence of the control
parameter mismatch of generators which have to be identical firstly
and the effect of nonlinear distortions in the communication channel
on the efficiency of the proposed and already known communication
schemes are also considered in this Section. Final discussions and
remarks are given in Conclusions.

\section{Theoretical background of the method to be proposed}
\label{sct:GSandNIS}

The generalized synchronization regime (GS) in two unidirectionally
coupled chaotic oscillators means the presence of a functional
relation $\mathbf{u}(t)=\mathbf{F}[\mathbf{x}(t)]$ between the drive
$\mathbf{x}(t)$ and response $\mathbf{u}(t)$ system
states~\cite{Rulkov:1995_GeneralSynchro,
Pyragas:1996_WeakAndStrongSynchro}. This relation may be rather
complicated and even fractal, with the form of this relation being
not usually found in most cases, that seems to be very important if
it is used for secure communication.

To detect the GS regime the conditional Lyapunov exponent
computation~\cite{Pyragas:1996_WeakAndStrongSynchro}, the nearest
neighbor~\cite{Pecora:1995_statistics} and auxiliary system
methods~\cite{Rulkov:1996_AuxiliarySystem} are frequently used. At
the same time, only the last of them can be easily realized in
practice. According to this method the behavior of the response
system $\mathbf{u}(t)$ is considered together with the auxiliary
system $\mathbf{v}(t)$ one. The auxiliary system is equivalent to
the response one by the control parameter values, but starts with
other initial conditions belonging to the same basin of chaotic
attractor (if there is the multistability in the system). In
practice it means the small distinctions in the initial conditions
that are realized automatically because of the presence of
fluctuations in the constructive elements of generators. If GS takes
place, the system states $\mathbf{u}(t)$ and $\mathbf{v}(t)$ become
identical after the transient is finished due to the existence of
the relations $\mathbf{u}(t)=\mathbf{F}[\mathbf{x}(t)]$ and
$\mathbf{v}(t)=\mathbf{F}[\mathbf{x}(t)]$. Thus, the coincidence of
the state vectors of the response and the auxiliary systems
$\mathbf{v}(t)\equiv\mathbf{u}(t)$ is considered as a criterion of
the GS regime presence.

Except for the complicated character of the functional relation
between the drive and response system states the GS regime has
several other advantages in comparison with the other types of
chaotic synchronization usually used for the secure information
transmission (e.g., the complete synchronization one). First, this
regime can be observed in the nonidentical and even different
dynamical systems (including the systems with the different
dimension of the phase space), that makes possible transmitting and
receiving generators to be nonidentical (see
Sections~\ref{sct:Method} and \ref{sct:Advantages} for detail).
Second, the location of the boundary of the GS regime onset on the
``parameter mismatch --- coupling strength'' plane differs radically
from the other synchronization types. In particular, there are known
examples of unidirectionally coupled dynamical systems for which the
coupling strength corresponding to the onset of the GS regime in the
case of the small parameter mismatch is twice as much as for the
same oscillators with parameters detuned
sufficiently~\cite{Zhigang:2000_GSversusPS,Harmov:2005_GSOnset_EPL}.
This peculiarity allows to provide the appearance and destruction of
the synchronous regime at small modulation of the control parameter
value that ensures the effective parameter modulation at the
information transmission. Third, the noise is known to do not affect
on the threshold of the GS regime onset, i.e. the GS regime appears
in unidirectionally coupled dynamical systems in the absence and
presence of noise for the same values of the coupling parameter
strength~\cite{Koronovskii:2006_CGLE_GS_JETP}. As a consequence,
noise may be used to provide the additional masking of the
information signal. Indeed, this regime has many similarities with
the noise-induced synchronization regime both in the mechanisms of
the synchronous regime arising and methods for its
detection~\cite{Hramov:2006_PLA_NIS_GS}.

All arguments mentioned above results in the conclusion that GS can
be effectively used for secure information transmission.

\section{A method for secure information transmission}
\label{sct:Method}

The scheme for secure information transmission based on the GS
phenomenon is shown in Fig.~\ref{fgr:TransScheme}. Information
signal $m(t)$ is encoded in the form of a binary code. One or more
control parameters of the chaos generator are modulated slightly by
the information binary signal in such a way that the characteristics
of the chaotic signal (i.e., its amplitude and frequency both in
time realization and power spectrum) are not changed noticeably. To
provide the additional masking of the information signal as well as
the variation of the characteristics of the transmitted signal the
source of noise is used.

The obtained signal is transmitted through the communication channel
(where the noise is supposed to be also observed) to the receiver
located on the other side of the communication channel. It
represents two identical chaos generators capable to be (or not to
be) in the GS regime with the transmitting one. The principle of
operation of the receiver is based on the GS regime detection by
means of the auxiliary system
method~\cite{Rulkov:1996_AuxiliarySystem}. At the output of the
receiver the transmitted signal passes on the subtractor and the
recovered signal $\tilde{m}(t)$ is detected.

The character of modulation of the control parameters of the
transmitting generator should be chosen in such a way that depending
on the transmitting binary bit ``0''/``1'' the existence or
nonexistence of the GS regime between the transmitter and receiver
chaos generators would be observed. In order the variation of the
characteristics of the transmitted signal remains unnoticeable for
the non-authorized third party, the location of the GS boundary on
the ``control parameter mistuning -- coupling parameter'' plane must
have the peculiarity discussed in Section~\ref{sct:GSandNIS}, i.e.
at the small variation of the control parameter the critical
coupling parameter value corresponding to the threshold of the GS
regime onset must change sharply. Then, if the GS regime is chosen
to take place when binary bit ``0'' is transmitted, due to the
existence of the functional relation between the chaotic system
states both generators of the receiver would demonstrate identical
oscillations. Therefore, upon passing the subtractor the absence of
chaotic oscillations, i.e. the binary bit ``0'', would be observed.
On the contrary, when the binary bit ``1'' is transmitted the GS
regime does not take place. Thus, the oscillations of the receiver
generators in this case would be non-identical. Upon passing the
subtractor the chaotic oscillations of non-zero amplitude, i.e. the
binary bit ``1'', would be detected.

It should be noted, that the particularity of the boundary of the GS
regime mentioned above provides interchange of regions with
synchronous and asynchronous behavior on the receiving side of the
communication channel depending on the transmitting binary bit at
negligible changes of the transmitting signal characteristics.
Furthermore, due to the presence of the noise source these changes
become completely unnoticeable, especially in the time
representation. That leaves non-authorized third party no
possibility to decode the message signal by the one transmitting
through the communication channel.

\section{Numerical simulation of the proposed method}
\label{sct:NumSim}

The efficiency of the proposed method can be verified with the
following numerical example. Transmitting and receiving generators
are chosen to be unidirectionally coupled R\"ossler systems. Such
choice is connected with the facts that (i) R\"ossler system is
studied in detail (including in the view of GS, see
e.g.~\cite{Zhigang:2000_GSversusPS,Hramov:2005_IGS_EuroPhysicsLetters,Aeh:2005_GS:ModifiedSystem,Harmov:2005_GSOnset_EPL}),
(ii) the location of the GS boundary fulfils the requirements given
in Sections~\ref{sct:GSandNIS}--\ref{sct:Method} (see
also~\cite{Harmov:2005_GSOnset_EPL}), (iii) it is possible to
construct electronic generator which dynamics is described by such
equations~\cite{Rico-Martinez:2003_RoesslerGenerator}.

The transmitting generator is given by:
\begin{equation}
\begin{array}{ll}
\dot x_1=-\omega_{x}x_2-x_3,\\
\dot x_2=\omega_{x}x_1+ax_2,\\
\dot x_3=p+x_3(x_1-c),\\
\end{array}
\label{eq:TransmitterRoesslers}
\end{equation}
where $\mathbf{x}(t)=(x_1,x_2,x_3)^T$ is its vector-state, $a=0.15$,
$p=0.2$ and $c=10$ are the control parameter values, $\omega_x$
defines the natural frequency of the system oscillations.

The control parameter $\omega_x$ is modulated by the message binary
signal in the following way. If in the given time interval the
binary bit ``1'' is transmitted, then the control parameter
$\omega_x=0.91$ in that time internal. At transmission of the binary
bit ``0'' parameter $\omega_x$ is chosen randomly in the range
$[0.9;0.9099]$ providing the relative maximal value of the
$\omega_x$-parameter mismatch to be less that 1.1\%. Such a choice
of the control parameter $\omega_x$ is caused by the character of
the GS boundary location studied in~\cite{Harmov:2005_GSOnset_EPL}.
It provides slight modulation of characteristics of the transmitting
signal both in time and frequency representations. In fact, we can
modulate this parameter quite arbitrary and even use modulation of
any or several parameters, but the key condition in this case is the
interchange of regions with the asynchronous dynamics and the GS
regime in the time series of the transmitted signal.

\begin{figure}[p]
\vspace*{5.0cm} \centerline{\includegraphics*[scale=0.7]{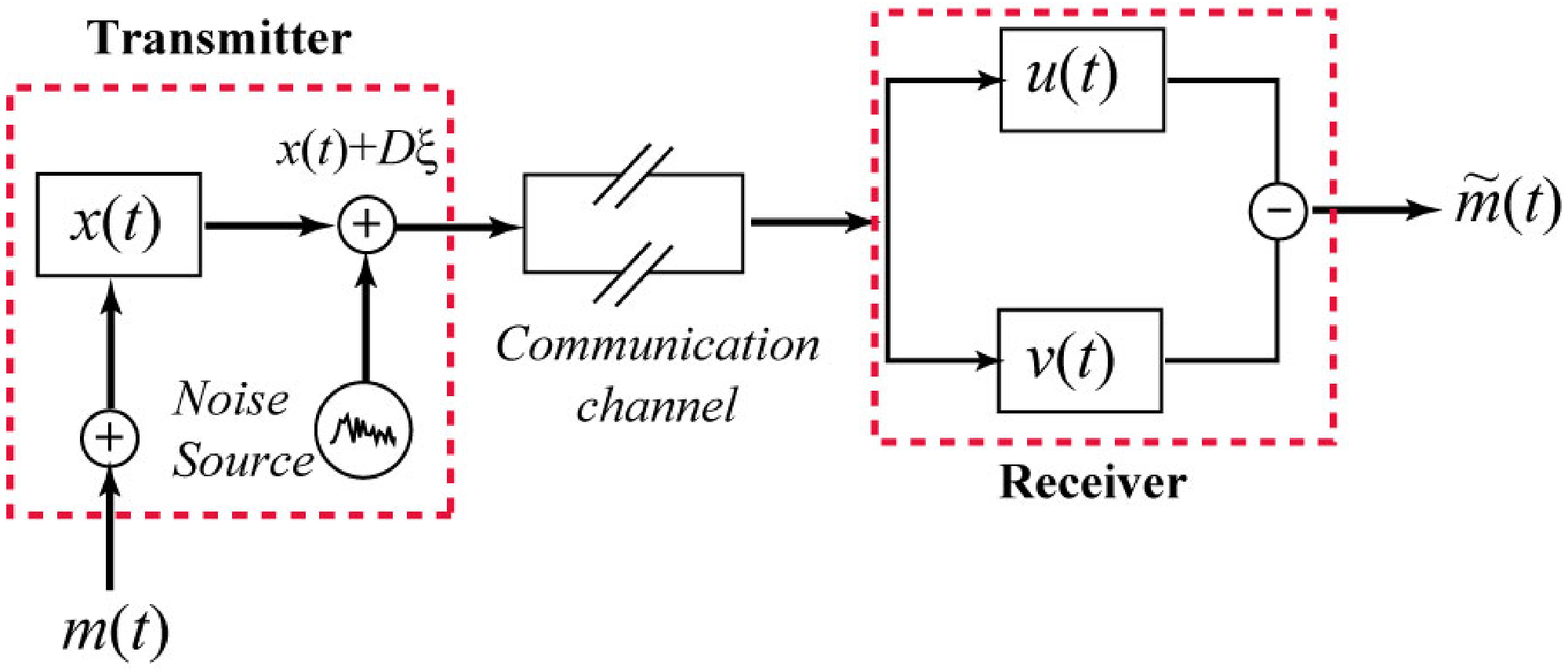}}
\caption{(Color online) The scheme for secure information
transmission based on GS. Here $m(t)$ is a binary information
signal, $x(t)$ is a vector-state of the transmitting generator,
$u(t)$ and $v(t)$ are of the receiving one, respectively;
$\tilde{m}(t)$ is a recovered signal \label{fgr:TransScheme}}
\vspace*{5.0cm}
\end{figure}

The receiver consists of two identical chaos generators each of
which is described by the following system:
\begin{equation}
\begin{array}{ll}
\dot u_1=-\omega_{u}u_2-u_3 +\varepsilon(s(t)-u_1),\\
\dot u_2=\omega_{u}u_1+au_2,\\
\dot u_3=p+u_3(u_1-c). \label{eq:ReceiverRoesslers}
\end{array}
\end{equation}
Here $\mathbf{u}(t)=(u_1,u_2,u_3)^T$ is the vector--state of the
first receiver generator. Let $\mathbf{v}(t)=(v_1,v_2,v_3)^T$, also
satisfying~(\ref{eq:ReceiverRoesslers}), be the vector-state of the
second generator (see Fig.~\ref{fgr:TransScheme}). Control
parameters $a$, $p$ and $c$ are selected identical to the
transmitter generator ones. The control parameter $\omega_u=0.95$
characterizing the natural frequency of the receiver generators is
chosen to be fixed for all time.

Signal from the transmitting generator after the summing with the
noise signal produced by the noise source (see
Fig.~\ref{fgr:TransScheme}) passes through the communication
channel. In our model it is realized by the coupling between the
transmitting and receiving generators, i.e. by adding the component
$\varepsilon(s(t)-u_1)$ in the first equation
of~(\ref{eq:ReceiverRoesslers}). Here $s(t)=x_1+D \xi$ is the
so-called signal of the communication channel\footnote{It should be
noted that the signal transmitting through the communication channel
comes to the receiver generators with certain delay. But due to
unidirectional coupling between the transmitting and receiving
generators this delay does not matter.}.  The term $D\xi$ simulates
the noise both appeared in the communication channel and produced by
the noise source. Here $\xi$ is the stochastic Gaussian process
which is described by the following probability distribution:
\begin{equation}
p(\xi)=\frac{1}{\sqrt{2\pi}\sigma}\exp\left(-\frac{(\xi-\xi_0)^2}{2\sigma^2}\right),
\label{eq:NormalDistrib}
\end{equation}
where $\xi_0=0$ and $\sigma=1.0$ are the mean value and
variance\footnote{It is important to note that the character of the
distribution of the random variable $\xi$ does not matter and the
similar results have been obtained both for uniform distribution of
the probability density $p(\xi)$ and Gaussian distribution with
different values of the mean value and variance.}. Parameter $D$
defines the total intensity of the noise.

The coupling strength between the transmitter and receiver
generators is characterized by the parameter $\varepsilon$. It is
chosen to be equal to $\varepsilon=0.14$. In that case, in the
absence of the noise ($D=0$) the GS regime
in~(\ref{eq:TransmitterRoesslers})-(\ref{eq:ReceiverRoesslers}) is
known to take place if $\omega_x<0.91$ or $\omega_x>0.97$, whereas
for $\omega_x\in[0.91;0.97]$ GS is not observed
(see~\cite{Harmov:2005_GSOnset_EPL} for detail).

The subtractor (see Fig.~\ref{fgr:TransScheme}) realizes operation
${(u_1-v_1)^2}$. Then upon passing it due to the auxiliary system
method the absence of oscillations for $\omega_x\in[0.9;0.9099]$ and
the presence of chaotic oscillations for $\omega_x=0.91$ should be
observed. The recovered signal $\tilde{m}(t)$ would represent the
sequence of regions with the different behavior.

\begin{figure}[p]
\centerline{\includegraphics*[scale=0.5]{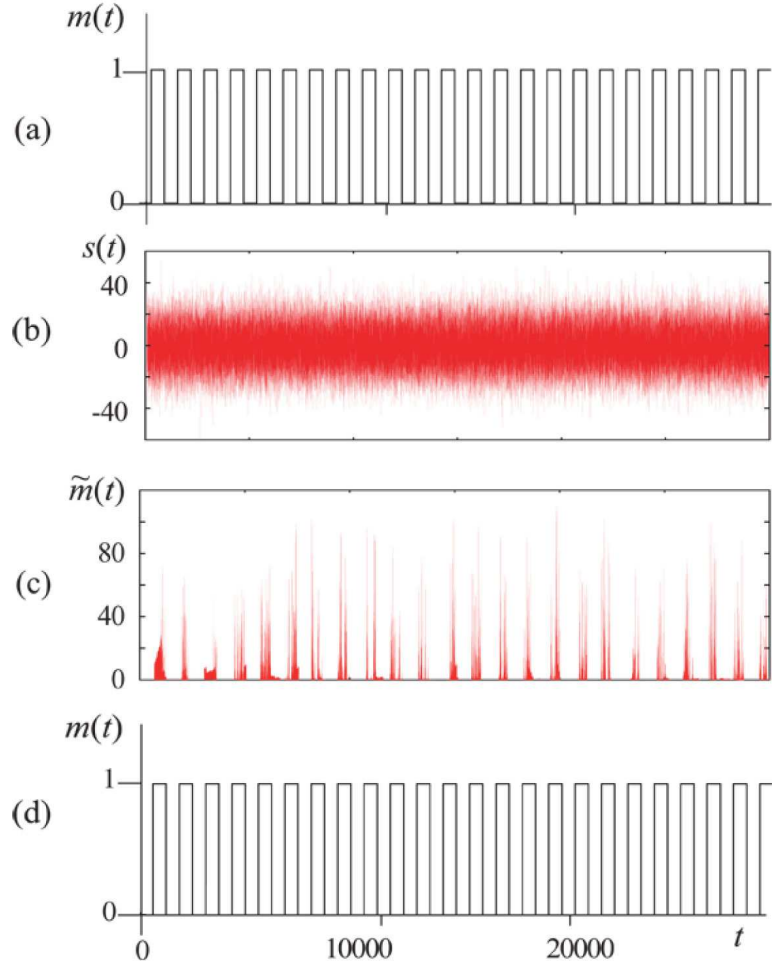}} \caption{(Color
online)(\textit{a}) Transmitted message signal $m(t)$ represented by
the sequence of the binary bits ``0''/``1'', (\textit{b}) the signal
$s(t)$ transmitted through the communication channel, (\textit{c})
recovered signal $\tilde{m}(t)=(u_1-v_1)^2$ and (\textit{d}) the
signal recovered by low-pass filtering and thresholding for value of
the noise amplitude $D=10$ \label{fgr:MessageSignal}}
\end{figure}

For the demonstrating purposes a simple sequence of the binary bits
``0''/``1'' presented in Fig.~\ref{fgr:MessageSignal},\textit{a} is
chosen to be a message signal $m(t)$. To integrate stochastic
equation~(\ref{eq:ReceiverRoesslers}) we have used Euler method with
time discretization step $\Delta t=0.0001$. It should be noted, that
the use of Euler method is not principal. In particular, the similar
results have also been obtained by us with the four order Runge-Kutt
method adapted for the stochastic differential
equations~\cite{Nikitin:1975_StochDE_eng}.

\begin{figure}[p]
\vspace*{5.0cm} \centerline{\includegraphics*[scale=0.6]{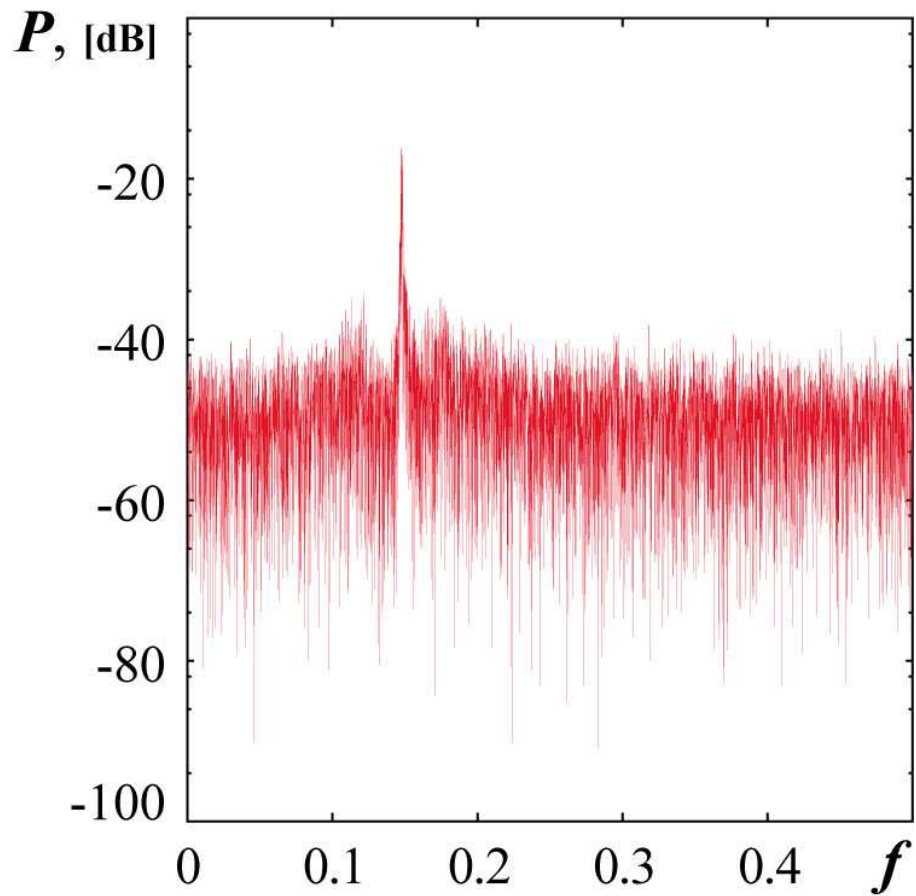}}
\caption{(Color online) Power spectrum of signal $s(t)$ transmitting
through the communication channel for the value of the noise
amplitude $D=10$. One can easily see only one narrow spectral
component in it. That makes decoding information signal difficult
without full information about characteristics of the receiver
\label{fgr:Spectrum}} \vspace*{5.0cm}
\end{figure}

Fig.~\ref{fgr:MessageSignal},\textit{b--d} illustrate the operation
of the scheme in the presence of noise. We have chosen the noise
intensity to be great enough, e.g. $D=10$. In
Fig.~\ref{fgr:MessageSignal},\textit{b} the signal transmitting
through the communication channel is shown. One can easily see that
modulation of the parameter $\omega_x$ does not change the
characteristics of the transmitting signal noticeably. Furthermore,
the noise of the great intensity still more distorts the
transmitting signal, first of all increasing its amplitude. Power
spectrum of such signal contains only one very narrow distinct
spectral component (see Fig.~\ref{fgr:Spectrum}), with it being
become almost indistinguishable with further increasing of the noise
intensity. In this case the non-authorized third party has no
possibility to decode the information signal without a full
information about characteristics of the receiver generators. In
particular, our calculations show that application of the windowed
Fourier transform or continuous wavelet
transform~\cite{alkor:2003_WVTBookEng} does not permit to detect
useful information from the transmitted signal.
Fig.~\ref{fgr:MessageSignal},\textit{c} shows the recovered signal
$\tilde{m}(t)=(u_1-v_1)^2$. The message signal is seen to be easily
recovered by the low-pass filtering and the thresholding the signal
$\tilde{m}(t)$. It is shown in Fig.~2,\textit{d}.

Other principal question connected with the secure communication
scheme operation is the rate of the information transmission. As we
have discussed in the Introduction, the rate of the proposed method
as well as other ones (transmitting digital information) based on
chaotic synchronization due to the presence of the transient
processes in switching is a low one. To illustrate this fact
numerically we estimate the rate of the information sending for our
scheme. It is clearly seen from Fig.~\ref{fgr:MessageSignal} that
for a time of 10000 units 18 bits of information have been
transmitted, i.e. the rate of the proposed method is
$1.8\cdot10^{-3}$ bits/unit of time.

The analogous results could be obtained for different values of the
noise intensity till $D=400$. Moreover, the recovered signals
$\tilde{m}(t)$ look qualitatively identical for different $D$, i.e.
the noise do not destruct the GS regime. In this case one can say
about the remarkable stability of our communication scheme to noise
despite of the fact that the stability of all schemes proposed
early~\cite{Cuomo:1993_ChaosCommunication,Dedieu:ChaoticSwitching1993,
Dmitriev:1995_ChaoticCommunications,Yand:ParamModul1996,Terry:GSchaosCom2001,LakshmananCompSign:1998}
is limited. At that, the rate of the information sending would be
the same low as for other schemes transmitting digital information
on the basis of chaotic synchronization.

\section{Stability of secure communication schemes to noise}
\label{sct:SNR} As it was mentioned in Introduction, we assume that
the stability of communication schemes to noise is one of the most
important features of secure communication schemes. To quantify the
degree of their stability the quantitative characteristics of their
efficiency in the presence of noise should be introduced. For
digital secure communication schemes such characteristics is an
average energy of chaotic radio pulse per transmitted information
bit $E_b$ related to the noise spectral density $N_0$, up to which
the secure communication scheme remains
efficient~\cite{Sklyar:CifrCommun2003Eng}. The energy per bit is
described by:
\begin{equation}
E_b={\rm P}_{sign}T, \label{eq:Energy_per_bit}
\end{equation}
where ${\rm P}_{sign}$ is the power of transmitted signal (without
noise), $T$ is a time spent for transmission of one bit of
information. The noise spectral density is defined as:
\begin{equation}
N_0=\frac{{\rm P}_{noise}}{B}, \label{eq:SpectrNoiseDens}
\end{equation}
where ${\rm P}_{noise}$ is a power of noise in the communication
channel, $B$ is the bandwidth of the channel.

The power of the signal (independently of the fact whether it is
deterministic or stochastic) has been computed by its time
realization. In our calculations the channel bandwidth has been
chosen to be ${B=f_2-f_1=0.2}$, where $f_1=0.05$, $f_2=0.25$ are its
boundary values.

To compare the effectiveness of our communication scheme in the
presence of noise with the several other ones we have estimated the
value of $E_b/N_0$, up to which secure communication scheme remains
efficient, for our scheme and a series of another schemes proposed
in earlier
publications~\cite{Cuomo:1993_ChaosCommunication,Dedieu:ChaoticSwitching1993,
Dmitriev:1995_ChaoticCommunications,Yand:ParamModul1996,Terry:GSchaosCom2001,LakshmananCompSign:1998}.
Certainly, we could not consider all schemes known at present but we
have tried to choose schemes most close to our one and touch upon
all well-known communication schemes which have already become
classical, i.e. chaotic masking, chaotic shift keying, nonlinear
mixing and others. In all cases the unidirectionally coupled
R\"ossler systems with the control parameter values being closed to
the values given in Section~\ref{sct:NumSim} have been chosen to be
a transmitter and receiver.

The $E_b/N_0$ values corresponding to the cases when there are no
possibility to recover the information signal as well as the names
of schemes and references to the papers, where they are described,
are shown in Table~\ref{tbl:Schemes}. It is clearly seen that all
considered schemes become fully inoperative for the positive values
of $E_b/N_0$, i.e. for the noise power less than the power of signal
transmitted through the communication channel. The radically
different situation takes place when our scheme is considered. Our
scheme remains efficient until $E_b/N_0$ is less than zero (see
Table~\ref{tbl:Schemes}). In the other words, our scheme possesses a
remarkable stability to noise. Even if the noise intensity
considerably exceeds the transmitting signal one, our scheme is
capable of working.

To explain such behavior of our communication scheme we should refer
to the following discussions. As we have already mentioned in
Section~\ref{sct:GSandNIS}, the GS regime on the basis of which our
communication scheme has been proposed, has many similarities with
the noise-induced synchronization one~\cite{Hramov:2006_PLA_NIS_GS},
i.e. we can achieve GS both by the deterministic and stochastic
signal influence. Due to the fact that both transmitter generators
have been affected by the influence of the same signal, the
character of signal (is it deterministic or stochastic one) is not
crucial. The key role belongs to the presence of the main spectral
components which are almost undistinguishable (see
Fig.~\ref{fgr:Spectrum}). When intensity of noise is such a great
one that spectral components disappear, our method become failed
because of the noise-induced synchronization regime realization.
Therefore our scheme possess a remarkable stability to noise.

A radically different situation takes place in the other schemes for
secure communication mentioned above. All of them demands the
presence of identical generators in the different sides of the
communication channel, at that in all of them the recovered signal
is obtained as a difference between signal in the communication
channel and response of the receiver generator on it. It is clear
that in the case of the influence of stochastic (random) signal on
the deterministic receiver generator the obtainment of the same
stochastic signal is impossible. Therefore for such secure
communication schemes their efficiency in the presence of noise is
limited by the amount of noise which the transmitter generator
itself could support. In the case of the use of R\"ossler generators
(with control parameter values closed to the last one presented in
Section~\ref{sct:NumSim}) in the transmitter and receiver such
amount in accordance with analogue of Shannon theorem for digital
signals~\cite{Sklyar:CifrCommun2003Eng} is $E_b/N_0=-1.46$~dB. It is
clearly seen that our numerical estimations are in a full agreement
with the theoretical prediction results.

\begin{figure}[p]
\vspace*{5cm} \centerline{\includegraphics[scale=0.6]{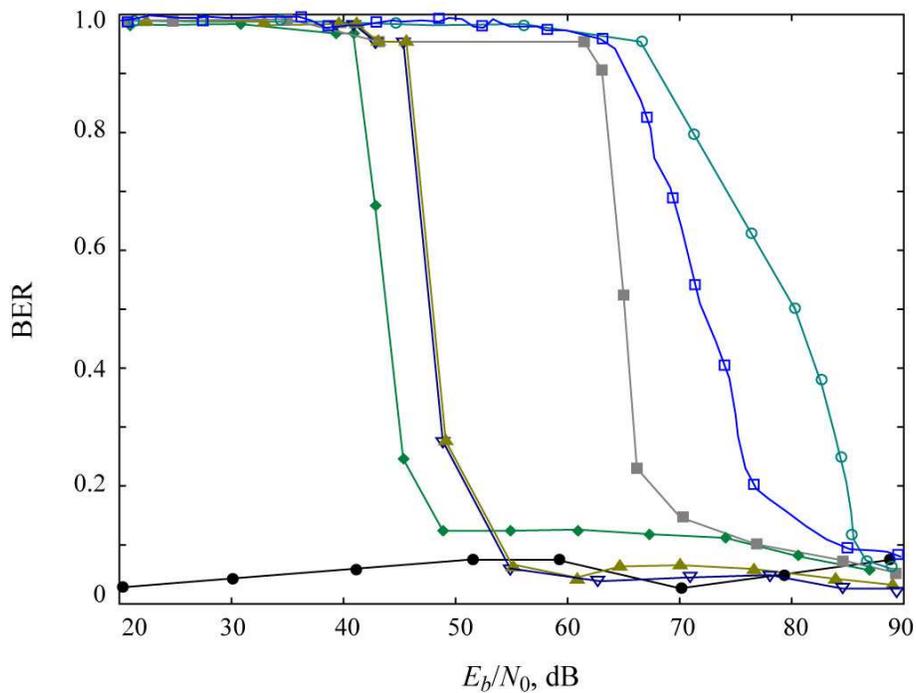}}
\caption{(Color online) Dependence of the bit error rate (BER) on
the average energy of chaotic radio pulse per transmitted
information bit related to the noise spectral density ($E_b/N_0$)
for different secure communication schemes, i.e. $\circ$ - chaotic
masking, $\blacklozenge$ - chaotic switching (chaotic parameter
modulation), $\blacksquare$ - nonlinear mixing, $\blacktriangle$ -
scheme on the basis of GS described in~\cite{Terry:GSchaosCom2001},
$\vartriangle$ - scheme on the basis of GS and CS described
in~\cite{Terry:GSchaosCom2001}, $\square$ - scheme with compound
signal~\cite{LakshmananCompSign:1998}, $\bullet$ - our secure
communication scheme\label{fgr:BER}} \vspace*{5cm}
\end{figure}

Correctness of the arguments mentioned above could be also confirmed
by the dependence of the bit error rate
(BER)~\cite{Abel:ChaosCom2002} on the $E_b/N_0$ value for different
communication schemes mentioned above. Such dependencies are shown
in Fig.~\ref{fgr:BER}. At computation of the bit error rate the
threshold value allowing to detect the original information binary
signal from the signal $\tilde{m}(t)$ has been chosen to be fixed
independently on the noise intensity affected on the transmitting
generator whereas it has been changed for the characteristics
presented in the Table~\ref{tbl:Schemes}. At the same time, it is
clearly seen that for different secure communication schemes (except
the our one) BER becomes equal to 1 quickly, whereas for our scheme
it is closed to 0 independently on the noise intensity. Such results
are in a good agreement with the last one presented above in
Table~\ref{tbl:Schemes}.

It should be noted, that the change of the control parameter values
and equations of generators could result in the variation of the
quantitative values of the average energy per bit related to the
noise spectral density, but the order of these magnitudes, relation
between them and the qualitative behavior of BER vs $E_b/N_0$ would
always remain the same.

\section{The other advantages of the proposed method}
\label{sct:Advantages} Except of the problem connecting with the
influence of noise in the communication channel, all schemes
considered in Section~\ref{sct:SNR} have some other disadvantages
and difficulties for practical purposes. The most part of schemes
(e.g., schemes 1--4 in Table~\ref{tbl:Schemes}) uses the complete
synchronization of chaotic
oscillators~\cite{Pecora:1990_ChaosSynchro,Pecora:1991_ChaosSynchro}
that, first of all, requires an identity of the transmitting and
receiving generators being located in the different sides of the
communication channel. It is a very big conceptual problem,
especially for a long time of operation of the devices. Second, in
some cases the parameters used in the transmitter can be estimated
from the transmitting signal, and, therefore, the information signal
can be extracted from it~\cite{Ponomarenko:2002_SignalExtracting}.

The scheme proposed in~\cite{Terry:GSchaosCom2001} (system 5 in
Table~\ref{tbl:Schemes}), as well as our one, is based on the
generalized synchronization. Despite of the fact that its stability
to noise is greater than in several other schemes (e.g., in schemes
1 and 3 from Table~\ref{tbl:Schemes}), the problem of the identity
of the generators on both sides of the communication channel still
remain unsolved. Furthermore, such scheme demands realization of the
complementary communication channel that produces additional
problems for practical realization.

In schemes 6 and 7 (see Table~\ref{tbl:Schemes}) both types of the
synchronous chaotic system behavior mentioned above, i.e the
generalized and complete synchronization, are used for the
information transmission. In the first case both of them are touched
in this process directly whereas in the second one generalized
synchronization is used for the creation of the compound signal to
be transmitted through the communication channel. Certainly, such
schemes are supposed to be more secure in comparison with the other
schemes mentioned above, but they demand the presence of additional
identical generators on different sides of the communication channel
and, in some cases, realization of the complementary communication
channel. Therefore, the practical realization of these schemes is a
very big problem. Moreover, the stability to noise, especially, for
the scheme 7, is a very low (see Table~\ref{tbl:Schemes}).

Contrary to all schemes mentioned above, our one does not require an
identity of chaotic generators on both sides of the communication
channel, and, therefore, it is a simple enough for the practical
realization.

Other important characteristic of the secure communication schemes
is their robustness to the parameter mismatch (PM). As it was
mentioned above, almost all known communication schemes demand the
presence of two or more identical generators on one or different
sides of the communication channel. Due to the problems connected
with the technical realization of such schemes the influence of the
control parameter mismatch of generators, that have to be identical
firstly, is a topical problem of the secure communication. To
realize the parameter mismatch in numerical simulation we should
replace any control parameter $A$ by $A(1\pm\eta)$ in one of
identical generators. Then $\eta$ would be the value of the
parameter mismatch (PM). We have estimated the relative values of
the $\omega_u$ parameter mismatch up to which the considered schemes
remain efficient. Since the choice of sign of the parameter mismatch
does not matter, we have used the sign $"+"$ in our calculations.
The obtained results are presented in Table~\ref{tbl:Schemes2}
(column 4) for all schemes considered above. It is easy to see, that
in that case our scheme has analogous. For example, the chaotic
switching and chaotic parameter modulation schemes (i.e., schemes 2
and 4 in Tables~\ref{tbl:Schemes} and \ref{tbl:Schemes2},
respectively) are capable of working if control parameter values are
detuned up to 2 {\footnotesize\%}. The degree of robustness of our
communication scheme to the parameter mismatch is the same. But in
our scheme such generators are located on the one side of the
communication channel, so their adjustment to identical state can be
easily realized.

Except the presence of noise, transmitting signal can undergo the
changes of another character in the communication channel. One of
such changes is the nonlinear distortions of the signal.
Transmitting signal $x(t)$ is most frequently supposed to have a
transformations in the form of cubic
nonlinearity~\cite{Dmitriev2002:ChaosCommunication}. So, after the
transmission through the communication channel the signal
$y(t)=x(t)(1-\alpha x^2(t))$, where $\alpha$ is a small enough,
would pass at the input of the receiver.

To estimate the influence of nonlinear distortions in the
communication channel the following quantitative characteristic has
been introduced:
\begin{equation}
{\rm ND}=10\lg\frac{P_x}{P_y}, \quad \rm{[dB].}
\end{equation}
Here $P_x$ and $P_y$ are the powers of signals $x(t)$ and $y(t)$,
respectively, computed by the time realization of the signal. In
that way
\begin{equation}
P_x=\int_0^T x^2(t)dt, \quad P_y=\int_0^T y^2(t)dt,
\label{eq:P_time}
\end{equation}
where $T$ is the time of the signal transmission. The quantity
mentioned above characterizes the maximal level of nonlinear
distortions for which secure communication scheme remains efficient.
The more there would be this value, the more the signal would be
distorted, therefore the more would be the maximal level of the
nonlinear distortions up to which the methods for secure information
transmission would still remain efficient.

The results of the influence of the nonlinear distortions in the
communication channel on the efficiency of the secure communication
schemes mentioned above are shown in Table~\ref{tbl:Schemes2}
(column 5). It is easy to see that our scheme proposed in
Sections~\ref{sct:Method} and \ref{sct:NumSim} exceeds all
considered analogues according to this characteristic. The maximal
level of the nonlinear distortions for it is equal to $\rm
ND=27.2$~dB. The most close characteristics correspond again to the
chaotic switching and chaotic parameter modulation schemes (schemes
2 and 4 in Table~\ref{tbl:Schemes2}). At the same time, they are
found to be rather less than the last one for our secure
communication scheme. Furthermore, both schemes mentioned above
possess the limited stability to noise whereas the stability of our
scheme is almost unrestricted for the real limits.

It is necessary to emphasize that the quantitative characteristics
of the efficiency of the secure communication schemes presented in
Tables~\ref{tbl:Schemes} and \ref{tbl:Schemes2} are obtained for
unidirectionally coupled R\"{o}ssler systems with control parameters
values equal or closed to the last ones described in
Section~\ref{sct:NumSim} and Gaussian distribution of noise in the
communication channel. The change of parameters and equations of
generators and characteristics of the noise source could result in
variation of the quantitative values of all considered
characteristics, but their order as well as the relation between
them would always remain the same. In particular, the similar
results have been obtained by us for
Chua~\cite{Chua:1986_DoubleScroll} and
Rulkov~\cite{Rulkov:1996_SynchroCircuits} generators used both the
transmitter and receiver and different kinds of the noise
distribution in the communication channel.

\section*{CONCLUSIONS}
\label{sct:Conclusions} In conclusion, we have developed a new
method for secure information transmission possessing a remarkable
stability to noise. It enables to use additional source of noise in
our communication scheme providing a big distortions of the
transmitted signal, with decoding the information signal by
non-authorized third party being become difficult. Moreover, it does
not require an identity of the chaotic generators on both sides of
the communication channel because of using the generalized
synchronization instead of the complete one. Therefore, it is simple
enough for the practical realization. Instead, the additional
receiver generator is used for the possibility of the generalized
synchronization regime detection. To demonstrate the principal
advantages of our method in comparison with the early developed ones
the signal to noise ratio and influence of the control parameter
mismatch of firstly identical generators and nonlinear distortions
in the communication channel are estimated numerically both for our
scheme and for the series of the other ones.

\section*{ACKNOWLEDGMENTS}
We thank Dr. Svetlana V. Eremina for the English language support.
We are grateful to the Referees of our paper for the useful comments
and remarks. This work has been supported by Russian Foundation for
Basic Research (projects 08-02-00102) and Federal special-purpose
programme ``Scientific and educational personnel of innovation
Russia''.



%
%
%

\newpage

\newpage

\begin{table}[p]
\caption{The values of an average energy of chaotic radio pulse per
transmitted information bit related to the noise spectral density
($E_b/N_0$, [dB]) corresponding to the cases when there are no
possibility to recover the information signal \label{tbl:Schemes}}
\begin{center}
\noindent
\begin{tabular}{|c |l |c |c|}
\hline No. & \textbf{Scheme} & \textbf{Ref.} & \textbf{$\bf{E_b/N_0},[dB]$}   \\
\hline 1 & Chaotic masking &\cite{Cuomo:1993_ChaosCommunication} &
56.48 \\
\hline 2 & Chaotic shift keying &\cite{Dedieu:ChaoticSwitching1993}
&
30.76 \\
\hline 3 & Nonlinear mixing
&\cite{Dmitriev:1995_ChaoticCommunications}
& 64.99 \\
\hline 4 & Chaotic parameter modulation &\cite{Yand:ParamModul1996}
&
30.76 \\
\hline 5 & GS scheme &\cite{Terry:GSchaosCom2001} & 23.66 \\
\hline 6 & Scheme based on CS and GS &\cite{Terry:GSchaosCom2001} &
39.52 \\
\hline 7 & Scheme with compound signal
&\cite{LakshmananCompSign:1998} &
39.24 \\
\hline 8 & Our scheme & & -10.01  \\
\hline
\end{tabular}
\end{center}
\end{table}


\begin{table}[p]
\caption{Maximal value of $\omega_u$ parameter mismatch (PM,
{\footnotesize \%}) and level of nonlinear distortions (ND, dB) in
the communication channel for which secure communication schemes
remain efficient \label{tbl:Schemes2}}
\begin{center}
\noindent
\begin{tabular}{|c |l |c |c|c|}
\hline No. & \textbf{Scheme} & \textbf{Ref.} & \textbf{PM} & \textbf{ND}  \\
\hline 1 & Chaotic masking &\cite{Cuomo:1993_ChaosCommunication} &
0.30 & 1.03\\
\hline 2 & Chaotic shift keying &\cite{Dedieu:ChaoticSwitching1993}
& 2.00 & 23.3\\
\hline 3 & Nonlinear mixing
&\cite{Dmitriev:1995_ChaoticCommunications}
& 0.30 & 0.26\\
\hline 4 & Chaotic parameter modulation &\cite{Yand:ParamModul1996}
& 2.00 & 23.3\\
\hline 5 & GS scheme &\cite{Terry:GSchaosCom2001} & 1.00 & 7.75\\
\hline 6 & Scheme based on CS and GS &\cite{Terry:GSchaosCom2001} & 0.50 & 4.83\\
\hline 7 & Scheme with compound signal
&\cite{LakshmananCompSign:1998} & 0.20 & 2.63\\
\hline 8 & Our scheme & & 2.00 & 27.2\\
\hline
\end{tabular}
\end{center}
\end{table}

\end{document}